\documentclass{aa}

\usepackage{graphicx}
\usepackage{upgreek}
\usepackage[colorlinks=true,linkcolor=red,urlcolor=black,citecolor=blue]{hyperref}

\usepackage{txfonts}
\graphicspath{{./figures/}}

\bibpunct{(}{)}{,}{a}{}{,}

\newcommand\exo{EXO~2030+375\xspace}

\newcommand\nustar{\textsl{NuSTAR}\xspace}

\newcommand\xmm{\textsl{XMM-Newton}\xspace}

\newcommand\rxte{\textsl{RXTE}\xspace}
\newcommand\nicer{{NICER}\xspace}

\newcommand\integral{\textsl{INTEGRAL}\xspace}
\newcommand\swift{\textsl{Swift}\xspace}
\newcommand\chandra{\textsl{Chandra}\xspace}
\newcommand\astrosat{\textsl{AstroSAT}\xspace}
\newcommand\isis{\texttt{isis}\xspace}
\bibpunct{(}{)}{;}{a}{}{,} 

\newcommand*{\pp}{\textcolor{black}}

\title{The giant outburst of \exo}
\subtitle{II: Broadband spectroscopy and evolution}

\titlerunning{Broadband spectroscopy of \exo}

 \author{R.~Ballhausen\inst{\ref{in:UMD}, \ref{in:GSFC}}
          \and P.~Thalhammer\inst{\ref{in:Remeis}}
          \and P.~Pradhan\inst{\ref{in:Embry},\ref{in:MIT}}
          \and E.~Sokolova-Lapa\inst{\ref{in:Remeis}}
          \and J.~Stierhof\inst{\ref{in:Remeis}}
          \and K.~Pottschmidt\inst{\ref{in:UMBC},\ref{in:GSFC}} 
          \and J.~Wilms\inst{\ref{in:Remeis}}
          \and J.B.~Coley\inst{\ref{in:Howard},\ref{in:GSFC}}
          \and P.~Kretschmar\inst{\ref{in:ESA}}
          \and F.~F{\"u}rst\inst{\ref{in:ESA}}
          \and P.~Becker\inst{\ref{in:GMU}}
          \and B.~West\inst{\ref{in:USNA}}
          \and{C.~Malacaria\inst{\ref{in:ISSI}}} 
          \and{M.T.~Wolff\inst{\ref{in:NRL}}}
          \and{R.~Rothschild\inst{\ref{in:UCSD}}}
          \and R.~Staubert\inst{\ref{in:IAAT}}
        }
   
         \institute{
           {University of Maryland College Park, Department of Astronomy, College Park, MD 20742, USA \label{in:UMD}}
         \and
           {CRESST and NASA Goddard Space Flight Center, Astrophysics Science Division, 8800 Greenbelt Road, Greenbelt, MD
           20771, USA \label{in:GSFC}}
         \and
           {Dr.\ Karl Remeis-Observatory \& ECAP, Universit\"at Erlangen-N\"urnberg, 96049
           Bamberg, Germany\label{in:Remeis}}
         \and
           {Department of Physics and Astronomy, Embry-Riddle Aeronautical University, 3700 Willow Creek Road,  Prescott, AZ 86301, USA\label{in:Embry}}
         \and
           {Massachusetts Institute of Technology, Kavli Institute for Astrophysics and Space Research, 70 Vassar St., Cambridge, MA 02139, USA\label{in:MIT}}
         \and
           {University of Maryland Baltimore County, 1000 Hilltop Circle, Baltimore, MD 21250, USA\label{in:UMBC}}
         \and
           {Department of Physics and Astronomy, Howard University, Washington, DC 20059, USA\label{in:Howard}}
         \and
           {ESA/ESAC, Camino Bajo del Castillo s/n, 28692, Villanueva de la Ca\~{n}ada, Madrid, Spain\label{in:ESA}}
         \and
           {George Mason University, Fairfax, VA 22030, USA\label{in:GMU}}
         \and 
          {Department of Physics, United States Naval Academy, Annapolis, MD 21402, USA \label{in:USNA}}
         \and
           {International Space Science Institute, Hallerstrasse 6, 3012 Bern, Switzerland \label{in:ISSI}} 
        \and
            {Space Science Division, U.S. Naval Research Laboratory, Washington, DC 20375--5352, USA\label{in:NRL}}
        \and
            {Department of Astronomy and Astrophysics, University of California San Diego, 9500 Gilman Dr., La Jolla, CA 92093, USA\label{in:UCSD}}
        \and
           {Institut für Astronomie und Astrophysik, Eberhard-Karls-Universit\"at T\"ubingen, Sand 1, 72076 Tübingen, Germany\label{in:IAAT}}
        }
        \date{RECEIVED: ACCEPTED:}
\begin{document}

\abstract{
In 2021, the high-mass X-ray binary \exo underwent a giant X-ray outburst, the
first since 2006, that reached a peak flux of ${\sim}600\,\mathrm{mCrab}$ (3--50\,keV).
The goal of this work is to study the spectral evolution over the course of the outburst, search for possible cyclotron resonance scattering features (CRSFs), and to associate spectral components with the emission pattern of the accretion column.
We used broadband spectra taken with the Nuclear Spectroscopic Telescope Array (\nustar), the Neutron Star
Interior Composition Explorer (\nicer), and \chandra near the peak and during the decline phase of the outburst. We describe the data with established empirical continuum models and perform pulse-phase-resolved spectroscopy. We compare the spectral evolution with pulse phase using a proposed geometrical emission model.
We find a significant spectral hardening toward lower luminosity, a behavior that is expected for super-critical sources. The continuum shape and evolution cannot be described by a simple power-law model with exponential cutoff; it requires additional absorption or emission components. We can confirm the presence of a narrow absorption feature at ${\sim}10\,\mathrm{keV}$ in both \nustar observations. The absence of harmonics puts into question the interpretation of this feature as a CRSF. The empirical spectral components cannot be directly associated with identified emission components from the accretion column.
%
}

\maketitle

\section{Introduction}
The Be X-ray binary \exo was discovered in 1985 during a strong X-ray outburst at a 2--30\,keV flux level of ${\sim}700$\,mCrab \citep{Parmar1985} and soon after identified as neutron star binary with a Be companion \citep{Motch1987}. In addition to its ${\sim}$42\,s rotation period, it is strongly variable on a wide range of timescales, including an ${\sim}200$\,mHz Quasi-Periodic-Oscillation (QPO) \citep{Angelini1989} and episodic flaring \citep{Klochkov2011}. \exo exhibits regular type~I outbursts associated with its periastron passage according to its ${\sim}46\,\mathrm{d}$ orbit \citep{Wilson2008}. Since its discovery, in addition to the 1985 outburst, only two more ``giant'' type II outbursts exceeding ${\sim}500$\,mCrab (3--50\,keV) have been observed, in 2006 and 2021. \exo has an extensive observing history that provided evidence for strong neutral and ionized absorption, soft emission components, and strong pulse-phase variability. Distance estimates are still uncertain. The extinction-based estimate of $7.1\pm0.2\,\mathrm{kpc}$ \citep{Wilson2002} was recently challenged by \textsl{Gaia} observations, which find a much shorter parallax distance of $2.4^{+0.5}_{-0.4}\,\mathrm{kpc}$ \citep{Bailer-Jones_2021}.

The presence of a CRSF for this source is still a source of
debate. The first indication for a CSRF was presented by
\citet{Reig1999}, which analyzed a series of Rossi X-ray Timing Explorer (\rxte) observations. These
authors found an absorption-like feature at
${\sim36}\,\mathrm{keV}$ in the combined pulse-phase-averaged High
Energy X-ray Timing Experiment (HEXTE) spectrum. The
pulse-phase-resolved spectra did not allow for the detection of this
feature. \citet{Reig1999} noted that they were unable to make a
definitive statement about the nature of this feature. A 10\,keV
absorption feature was first detected in International Gamma-Ray Astrophysics Laboratory (\integral) observations of the
2006 outburst by \citet{Klochkov2007}, who also found indications of a
harmonic feature around 20\,keV, in principle supporting an
interpretation as a CRSF. However, \citeauthor{Klochkov2007} also
presented an equally acceptable description of the spectrum with a broad
emission line at ${\sim}13\text{--}15\,\mathrm{keV}$ instead of two
absorption features and cautioned that a CRSF detection is by no means
conclusive. \citet{Wilson2008} confirmed the presence of an absorption
feature near 10\,keV in \rxte observations of the 2006 giant
outburst. Most recently, \citet{Tamang2022} presented an analysis of
Nuclear Spectroscopic Telescope Array (\nustar) data also used in this work that further supports the presence of this feature. \citet{Klochkov2008} revisited \integral data of the 2006 giant outburst and found indications of an absorption line at ${\sim}63\,\mathrm{keV}$. This feature, however, only appears in a narrow pulse phase interval.

\exo has been monitored extensively at various luminosity stages and with all major X-ray missions. The evolution of the broadband spectrum has been presented by \citet{Wilson2008} and \citet{Epili2017} and has been extended down to the propeller regime by \citet{Fuerst2017} and \citet{Jaisawal2021}. In 2021 July, \exo went into its third and most recent giant X-ray outburst. In this paper, we focus on an in-depth spectral analysis of the \nustar data taken near the peak in the decline phase of the outburst, connecting our results to the characteristic pulse profile evolution presented by \citeauthor{Thalhammer2022} (\citeyear{Thalhammer2022}; hereafter Paper~I).

The remainder of the paper is structured as follows. In
Sect.~\ref{sec:data-red}, we describe the data acquisition with the
Neutron Star Interior Composition Explorer (\nicer) and \nustar over the course of the outburst and give details of the data reprocessing. %
In Sects.~\ref{sec:phaseavg} and~\ref{sec:phaseres}, we provide detailed pulse-phase-averaged and pulse-phase-resolved spectral analyses. We discuss our results in Sect.~\ref{sec:discussion} and give an outlook in Sect.~\ref{sec:outlook}.

\section{Data acquisition and reduction}\label{sec:data-red}

\subsection{Monitoring of the 2021 giant outburst of \exo}

\begin{figure}
    \resizebox{\hsize}{!}{\includegraphics{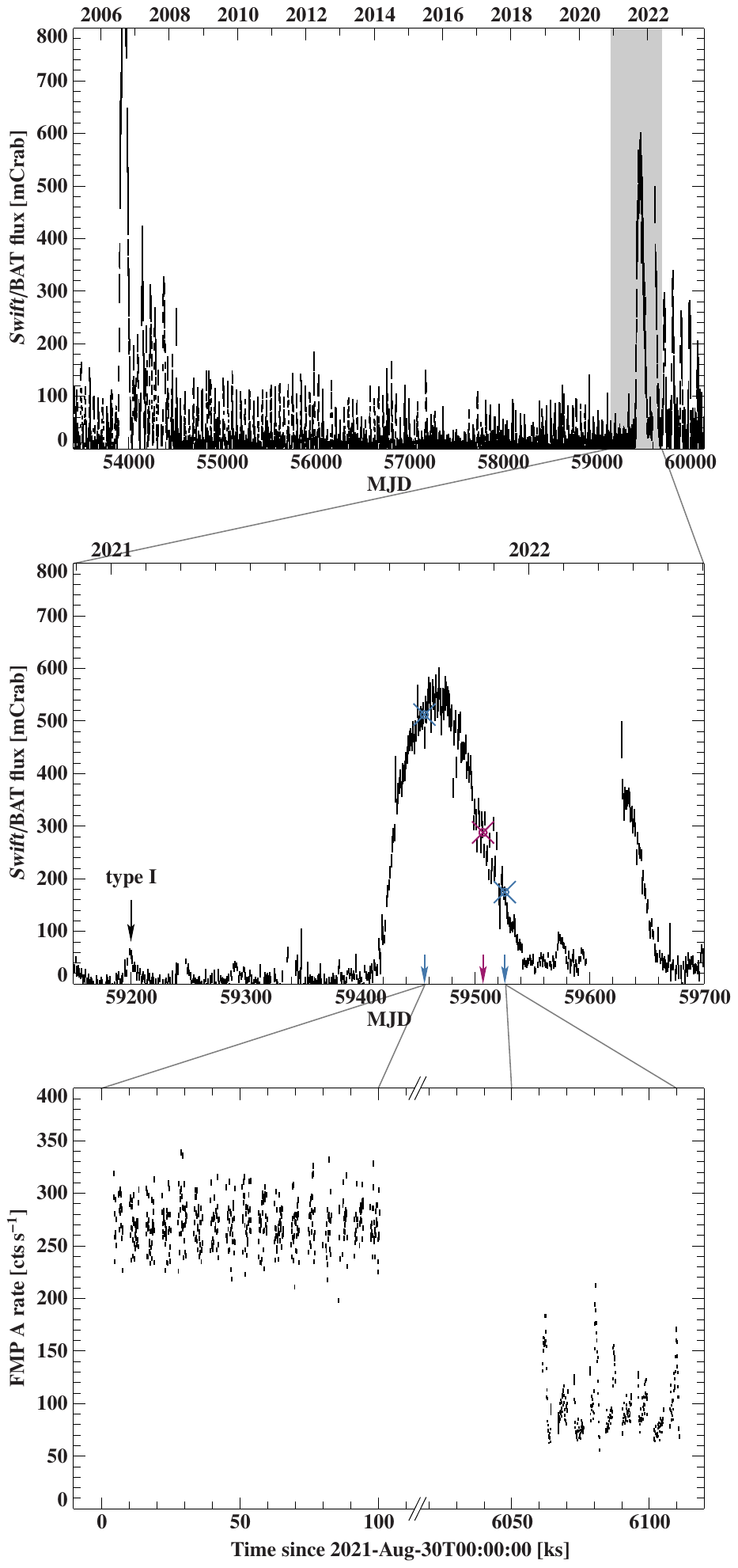}}
    \caption{Variability of \exo on different timescales. \textsl{Top:} Long-term \swift/BAT light curve of the giant 2022 outburst of \exo \citep{Krimm2013}.  \textsl{Middle:} Close-up of the 2021 giant outburst.  Blue crosses mark the times of the \nustar observations, and the magenta cross marks the time of the \chandra observation. \textsl{Bottom:} \nustar 3--78\,keV light curve of Obs.~I and II with 100\,s time resolution.}
    \label{fig:batlc}
\end{figure}

The 2021 giant outburst of \exo was closely monitored with \nicer and \swift/XRT (see Fig.~\ref{fig:batlc} for the \swift/BAT light curve). The monitoring campaign was complemented by pointed \nustar observations, once near the peak at a 3--50\,keV flux of ${\sim}600$\,mCrab (ObsID: 80701320002; hereafter Obs.~I) and during the decline phase around 250\,mCrab (ObsID: 9070133600; hereafter Obs.~II). The scheduling of the \nustar observations was motivated by a characteristic transition in the pulse profile and hardness ratio that had been reported in previous outbursts and was again observed through the dense \nicer and \swift/XRT monitoring \citep[Paper~I, ][]{Thalhammer2021a,Pradhan2021a}. Furthermore, the 2021 giant outburst of \exo was observed by \chandra, \astrosat, and \textsl{Insight}-HXMT \citep{Fu2023}.

\begin{table}
    \centering
    \label{tab:obs-log}
    \caption{Log of all observations used in this work.}
    \begin{tabular}{ccc}
    \hline\hline
    ObsID & start date & net exposure [ks]\\
    \hline
    \multicolumn{3}{c}{\textbf{\nustar}} \\
    \hline
    80701320002 & 2021 August 30 & 32.5\tablefootmark{a} \\
    90701336002 & 2021 November 08 & 23.5\tablefootmark{a} \\
    \hline
    \multicolumn{3}{c}{\textbf{\nicer}} \\
    \hline
    4201960113 & 2021 August 30 & \phantom{0}2.3 \\
    4201960150 & 2021 November 08 & 15.4 \\
    \hline
    \multicolumn{3}{c}{\textbf{\chandra}} \\
    \hline
    26154 & 2021 October 20 & \phantom{0}9.7\tablefootmark{b}\\
    \hline\hline
    \end{tabular}
    \tablefoot{
        \tablefoottext{a}{Per focal plane module.}
        \tablefoottext{b}{Medium energy grating  order +1.}
        }
\end{table}

\subsection{\nustar}

The NASA Small Explorer Mission \nustar \citep{Harrison2013}, launched on 2012 June 13, carries two co-aligned grazing incidence X-ray telescopes, which focus X-rays on the two focal plane modules, FPMA and FPMB. It provides a usable energy range of 3--79\,keV with a field of view of 13\arcmin$\times$13\arcmin at 10\,keV. \nustar's imaging capabilities allow for simultaneous background determination, while the pixelated CdZnTe detectors do not suffer from pile-up for count rates up to $10^5 \,\mathrm{cts}\,\mathrm{s}^{-1}\,\mathrm{pixel}^{-1}$.

We reprocessed the data of both \nustar observations with the standard \texttt{NUSTARDAS} pipeline version 2.0.0 with CalDB version 20211020. The source and background regions for both observations are circles of 90\arcsec radii, centered on the source and in the opposite corner of the field of view, respectively, to avoid contamination with source photons. We note that even at around 60\,keV the source count rate is still about an order above the background rate, so systematic effects of measuring the background on a different chip are expected to be negligible. Our screening results in ${\sim}32\,\mathrm{ks}$ cleaned exposure per FPM for Obs.~I and ${\sim}23\,\mathrm{ks}$ for Obs.~II. Event times were converted to the solar barycenter.
 
\subsection{\nicer}

The Neutron star Interior Composition Explorer \citep[\nicer;][]{Gendreau2016} was installed on the International Space Station in 2017. Its X-ray Timing Instrument (XTI) consists of 56 X-ray ``concentrator'' optics in combination with silicon detectors. Unlike \nustar, the XTI is not an imaging instrument, so background rates have to be reconstructed from models. \nicer features a timing resolution of $300\,\mathrm{ns}$ with a large collecting area over a bandpass of 0.2--12\,keV.

Out of the extensive \nicer monitoring data, which was analyzed and presented in detail in Paper I, we only used observations contemporaneous with \nustar to extend soft coverage (ObsID 4201960113 with \nustar Obs.~I with a net exposure of ${\sim}2.3\,\mathrm{ks}$ and ObsID 4201960150 for \nustar Obs.~II with a net exposure of ${\sim}15.4\,\mathrm{ks}$). Data reduction follows standard procedures using the \nicer data analysis pipeline of Heasoft version 6.31.1 with CALDB version 20220413. We used the \texttt{SCORPEON} background model\footnote{\url{https://heasarc.gsfc.nasa.gov/docs/nicer/analysis_threads/scorpeon-overview/}}. A detailed description of the data reduction is given in Paper I.

 \subsection{Chandra}

The Chandra X-Ray Observatory consists of the High Energy Transmission Grating Spectrometer (HETG) with a medium-energy grating (MEG; 0.4--5.0\,keV), and a high-energy grating (HEG; 0.8--10.0\,keV). The HETG have resolving powers up to about 1000 and effective area up to about $150\,\mathrm{cm}^2$. The dispersed grating spectra are recorded with an array of CCDs (Advanced CCD Imaging Spectrometer, ACIS-S; \citealt{Garmire2003}). We requested DDT observations of the source near the peak of the outburst, but owing to a problem with the insertion of the low-energy grating around that time, we obtained grating observations of the source for 10\,ks in FAINT TIMED mode only around 2021 October 20 (ObSID 26154), when the source was already in decline. The count spectra for the observations were extracted using
\texttt{TGCat} reprocessing scripts \citep{Huenemoerder2011} and using
CIAO \citep{CIAO:2006} version 4.8 and the Calibration Database version
4.7.2.

\section{Pulse-phase-averaged spectroscopy}\label{sec:phaseavg}

We examined the \nustar light curves in various energy bands over the course of the respective observations. The light curve of Obs.~I is rather stable, while Obs.~II exhibits strong flaring activity on kilosecond timescales (see Fig.~\ref{fig:batlc}). If we compare the variability of the hardness ratios over a range of energy bands with the light curves, however, they appear remarkably constant in both observations, indicating no significant spectral variability over the course of the individual observations. This behavior allows us to begin our spectral analysis from time- and pulse-phase-averaged spectra, where we expect the spectral shape to be only dependent on the luminosity. 

\begin{figure}
    \centering
    \resizebox{\hsize}{!}{\includegraphics{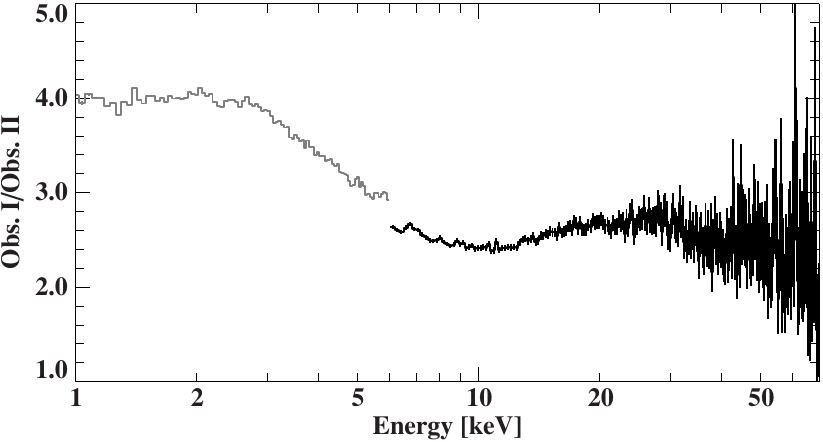}}
    \caption{Energy-resolved count-rate ratio of Obs.~I over Obs.~II. The gray curve is \nicer/XTI; the black one is \nustar/FPMA. The offset between \nicer and \nustar points at flux cross-calibration uncertainties between the instruments.}
    \label{fig:obsratio}
\end{figure}

In a first step, we investigated the spectral change from one observation to the other in a model-independent way by calculating energy-resolved ratios of the count rates of both observations in \nicer and \nustar (see Fig.~\ref{fig:obsratio}). We find Obs.~I to be significantly softer than Obs.~II, with a prominent excess of emission mostly below ${\sim}5\,\mathrm{keV}$. However, we note that the spectral evolution from Obs.~I to Obs.~II is more complex than can be fully captured by spectral hardness, and distinct differences in spectral shape become apparent over the full \nustar energy range.

We used \isis v.1.6.2-51 \citep{Houck2000a} for spectral fitting. 
We rebinned the data according to the optimal binning scheme of \citet{Kaastra2016}, but we required at least 25 counts per spectral bin.  We note a significant mismatch between \nicer and \nustar between 6\,keV and 10\,keV, which is likely due to cross-calibration issues, so we used an energy range of 1--6\,keV for \nicer and 4.5--78\,keV for \nustar. We further added a 1\% systematic uncertainty for the \nicer data to account for possible calibration uncertainties\footnote{See \nicer calibration memo NICER-Cal-SysErr-20221001 which estimates ${\sim}1\%$ systematic errors dominated by the response matrix for bright targets.}. Our model includes constants for flux cross-calibration uncertainties relative to \nustar/FPMA, and we fitted a gain-shift offset for all three instruments to account for energy calibration uncertainties around astrophysical and instrumental absorption edges. We find that permitting a slight gain shift leads to significant improvements of the fit statistic of $\Delta\chi^2\sim40$ for Obs.~I and $\Delta\chi^2\sim25$ for Obs.~II (for the respective best-fit model). The gain shifts we obtained are within 1--2 RMF energy bins (\nicer/XTI: 10\,eV and \nustar/FPM: 40\,eV)\footnote{The estimated uncertainties of \nustar's and \nicer's energy calibration are ${\sim}40\,\mathrm{eV}$ \citep{NuSTAR2022} and a ${\sim}5\,\mathrm{eV}$ (see NICER-Cal-Energy-Scale-optmv13-20221001), respectively.}. Uncertainties are given at the 90\% confidence level unless otherwise noted and obtained by varying a single parameter of interest for a $\Delta\chi^2=2.71$.

Our continuum model is based on earlier analyses of \exo \citep[e.g.,][]{Klochkov2011} and consists of an absorbed power law with exponential folding and a black-body component. We used the abundances for the interstellar medium from \citet{Wilms2000} and the cross-sections of \citet{Verner1996} to model neutral photoelectric absorption. Also based on previous observations \citep[e.g.,][]{Naik2013,Ferrigno2016}, 
we include a partial covering absorber distribution in order to model absorption in the binary system. The Fe K$\alpha$ complex is described by a broad Gaussian emission line in \nustar Obs.~I and II. 
 

 \begin{table}
 \centering
 \caption{Best-fit parameters for phase-averaged spectrum of \nustar Obs.~I and II. A partial covering absorber is not required in Obs.~II.}
 \label{tab:phasavg_bestfit}
    \begin{tabular}{lcc}
      \hline\hline
      parameter & Obs.\ I & Obs.\ II \\
      \hline
      $c_\mathrm{FPMB}$ & $0.9878^{+0.0018}_{-0.0017}$ &  $0.9988^{+0.0030}_{-0.0029}$ \\
      $c_\mathrm{XTI}$ &  $1.032\pm0.006$ & $0.939\pm0.005$ \\
       $N_{\mathrm{H},1}$\tablefootmark{a} & $3.9^{+0.8}_{-0.6}$
                          & $2.64\pm0.03$ \\
      $N_{\mathrm{H},2}$\tablefootmark{a} & $2.15^{+0.25}_{-0.47}$ & --- \\
      pcf &  $0.71^{+0.18}_{-0.16}$ &  --- \\
      $kT_\mathrm{BB}$\tablefootmark{b} & $0.472^{+0.036}_{-0.028}$ & $2.44\pm0.04$ \\
      $F_\mathrm{BB}$\tablefootmark{c} &  $0.58^{+0.18}_{-0.13}$ & $0.47\pm{0.03}$  \\
      $\Gamma$ & $1.114^{+0.011}_{-0.012}$ & $1.019\pm0.014$  \\
      $E_\mathrm{fold}$\tablefootmark{b} & $18.22^{+0.17}_{-0.18}$ &  $19.3\pm0.4$ \\
      $F_\mathrm{PL}$\tablefootmark{d} & $15.51^{+0.08}_{-0.07}$ & $4.81^{+0.04}_{-0.05}$ \\
      $A_\mathrm{Fe}$\tablefootmark{e} &  $6.7\pm0.4$ &  $1.91^{+0.23}_{-0.21}$  \\
      $E_\mathrm{Fe}$\tablefootmark{b} &  $6.594^{+0.023}_{-0.019}$ & $6.54\pm0.05$ \\
      $\sigma_\mathrm{Fe}$\tablefootmark{b} & $0.277^{+0.018}_{-0.017}$ &
                                                         $0.33\pm0.05$
      \\
      $E_\mathrm{10\,keV}$\tablefootmark{b} & $10.23^{+0.13}_{-0.15}$ & $9.4\pm0.4$ \\
      $\sigma_\mathrm{10\,keV}$\tablefootmark{b} & $2.58^{+0.29}_{-0.23}$ & $1.8^{+0.6}_{-0.4}$\\
      $d_\mathrm{10\,keV}$\tablefootmark{b} &  $0.32^{+0.06}_{-0.05}$ &  $0.13^{+0.10}_{-0.05}$ \\
      $\mathcal{L}$\tablefootmark{f} & $1.1\times10^{37}$ &  $3.6\times10^{36}$\\
      $\mathrm{GS}_\mathrm{FPMA}$\tablefootmark{g} &  $-29^{+15}_{-19}$ &  $-80^{+29}_{-20}$ \\
      $\mathrm{GS}_\mathrm{FPMB}$\tablefootmark{g} & $-37^{+14}_{-18}$ & $-75^{+27}_{-25}$ \\ 
      $\mathrm{GS}_\mathrm{XTI}$\tablefootmark{g} & $-9\pm4$ & $1.2^{+2.8}_{-2.7}$\\
      \hline\hline
      \end{tabular}
      \tablefoot{
        \tablefoottext{a}{In $10^{22}\,\mathrm{cm}^{-2}$.}
        \tablefoottext{b}{In keV.}
        \tablefoottext{c}{Unabsorbed 1--10\,keV flux in $10^{-9}\,\mathrm{erg}\,\mathrm{s}^{-1}\,
        \mathrm{cm}^{-2}$.}
        \tablefoottext{d}{Unabsorbed 3--50\,keV flux in $10^{-9}\,\mathrm{erg}\,\mathrm{s}^{-1}\,
        \mathrm{cm}^{-2}$.}
        \tablefoottext{e}{Photon flux in $10^{-3}\,\mathrm{phs}\,\mathrm{cm}^{-2}\,\mathrm{s}^{-1}$.}
         \tablefoottext{f}{3--50\,keV luminosity in $\mathrm{erg}\,\mathrm{s}^{-1}$ for a distance of 2.4\,kpc.}   
        \tablefoottext{g}{Additive gain shift in eV.}
      }
  \end{table} 

\begin{figure*}
\resizebox{\hsize}{!}{\includegraphics{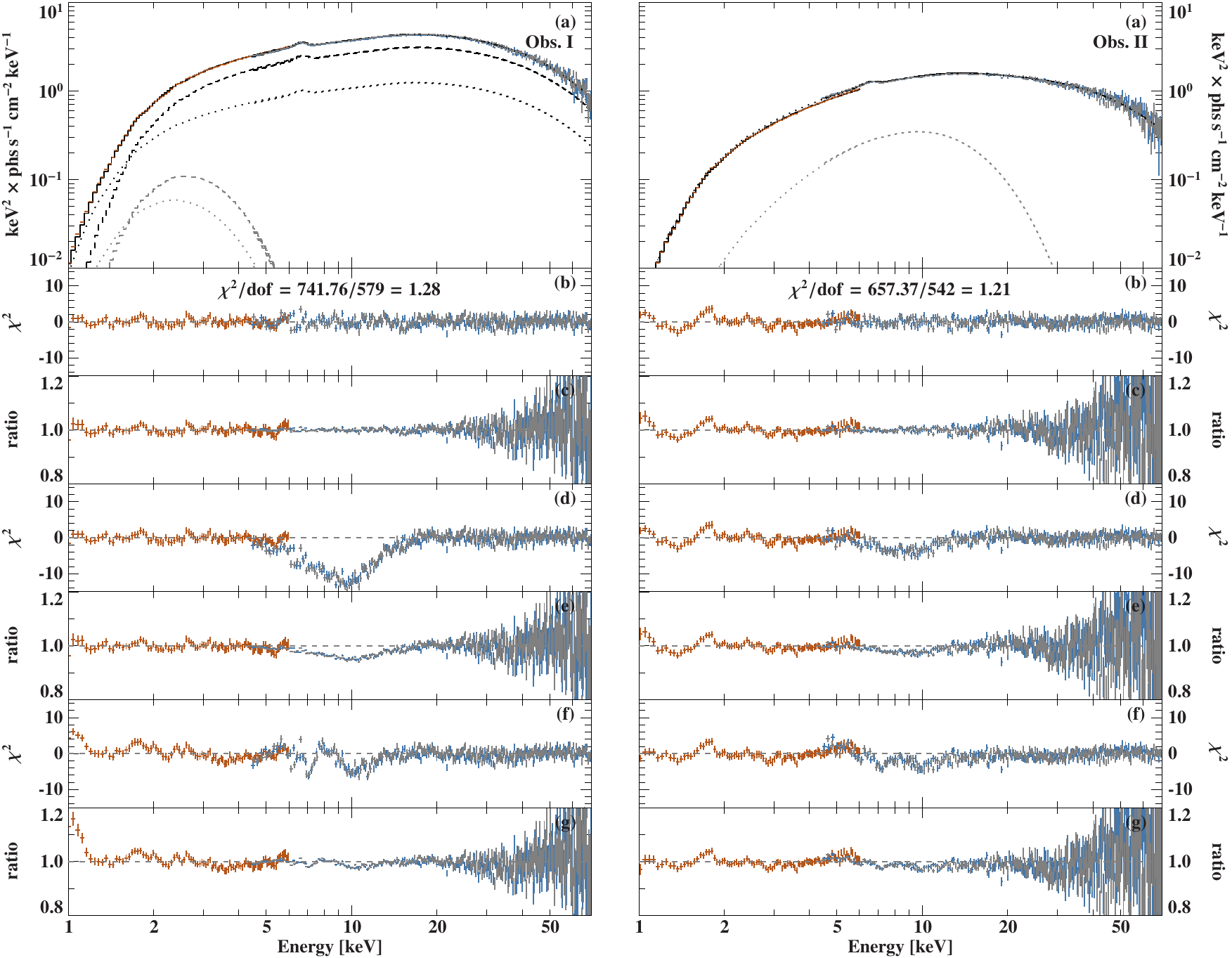}}
 \caption{Broadband spectra of Obs.~I and II. Panel a: Pulse-phase-averaged $\nu F_\nu$ spectrum Obs.~I (\textsl{left}) and Obs.~II (\textsl{right}) of \exo with best-fit, absorbed-cutoff power-law model for \nustar/FPMA (gray), FPMB (blue), and \nicer/XTI (orange). Panels b and c: Residuals and ratio for the best-fit model. The black dashed and dotted lines show the individual partial covering continuum components. The gray dashed and dotted lines show the partially absorbed black-body component only. Panels d and e: Residuals and ratio for the model evaluated without the 10\,keV absorption feature. Panels f and g: Residuals and ratio for the model re-fit without the 10\,keV absorption feature.}
 \label{fig:phsavg_obsI+II}
\end{figure*}

We found the inclusion of the 10\,keV absorption feature (\texttt{gabs} in \texttt{isis/xspec}; absorption line with Gaussian optical depth) mentioned above to result in the largest improvement in $\chi^2$, namely from 1256.32 to 741.76 for Obs.~I, and 713.14 to 657.37 for Obs.~II, while the $\chi^2$ improvement for the 36\,keV feature in Obs.~I was only $\Delta\chi^2=11.92$. We note that the temperature of the black-body component increases by roughly a factor of five from Obs.~I to Obs.~II. This component appears to be a broad contribution to the moderately hard emission resembling the ``10-keV-feature'' seen in emission in several other pulsars \citep[e.g., 4U\,0115+63][]{}. The black-body emission component in Obs.~II peaks roughly at the location of the 10\,keV absorption feature. The best-fit parameters are listed in Table~\ref{tab:phasavg_bestfit}, and Fig.~\ref{fig:phsavg_obsI+II} shows the respective spectra with best-fit model and residuals.  In \texttt{isis}/\texttt{xspec} notation, our full model is
\begin{equation*}
   S(E)= \texttt{detconst}\times\texttt{tbabs}\times(\texttt{bbody}+\texttt{cutoffpl}+\texttt{egauss})*\texttt{gabs}~,
\end{equation*}
 where for Obs.~II the partial covering fraction is unity as for a homogeneous absorber.

Interestingly, if line-like absorption features are included in Obs.~II, a partial covering absorber does not significantly improve the fit. We also note that the requirement of a partial covering absorber in Obs.~I is mostly based on a soft excess below ${\sim}2\,\mathrm{keV}$ in \nicer/XTI, which may be subject to further calibration uncertainties or inadequate background modeling or unaccounted-for emission components in our current model. We therefore conclude that there is no strong support for a partial covering absorber from broadband continuum absorption or edge absorption, and other possible interpretations include a partially ionized absorber that cannot be resolved with \nicer and \nustar. The Fe K emission complex indicates a certain degree of ionization due to its width and centroid energy. Our best-fit model therefore only includes a partial absorber component for Obs.~I.

\pp{For the spectral fitting of the \chandra/HETG data, we combined orders $+1$ and $-1$ of MEG and HEG, respectively, and rebinned the data to signal-to-noise ratios of 25, 5, and 10 in the respective energy bands 1--6\,keV, 6--7\,keV, and 7--10\,keV. We only fit the 1--10\,keV spectrum and used $C$-statistics \citep{Cash1979} for this dataset because of its lower statistic and our focus on weak line features. Due to the lack of hard X-ray coverage at the time of the \chandra observation, we find a continuum model used in Paper~I, consisting of an absorbed power law with an additional black body, sufficient. The pure continuum model resulted in a best-fit statistic of $C$-stat/d.o.f.\ of $542.82/426=1.27$.}

\begin{figure}
    \centering
    \resizebox{\hsize}{!}{\includegraphics{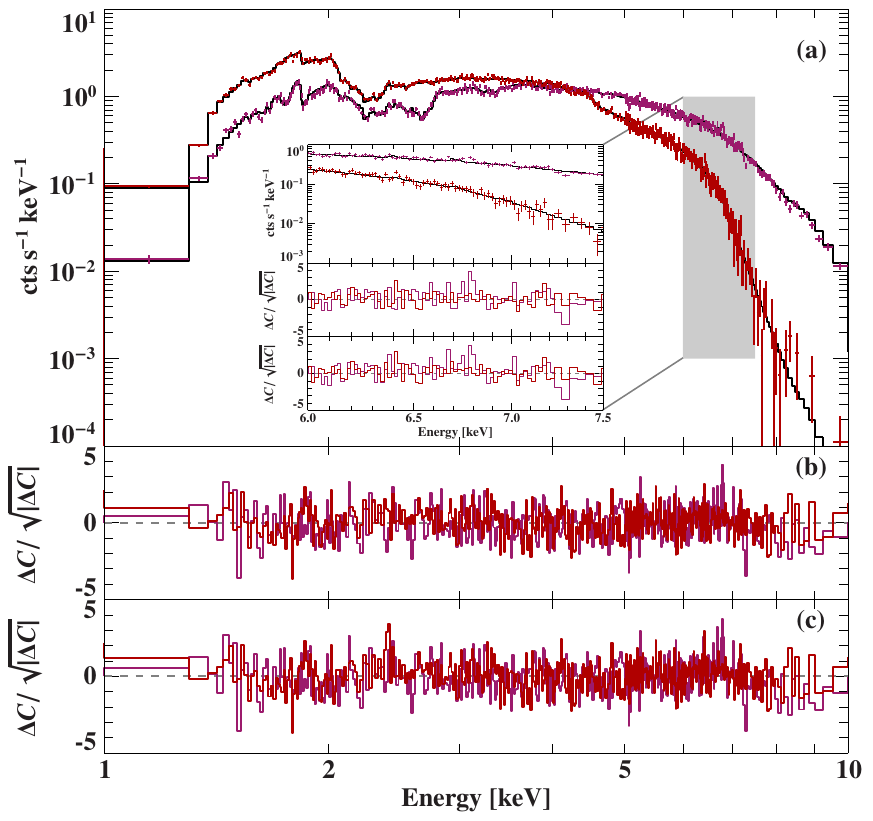}}
    \caption{\chandra/HETG spectrum of \exo. Panel (a) shows the count spectrum of MEG (red) and HEG (magenta) with the best-fit model (black). The inset shows a close-up of the iron band. Panel (b) shows $C$ residuals for the best-fit model with \ion{Si}{xiv}\,Ly$\alpha$, Fe\,K$\alpha$, \ion{Fe}{xxv}\,He$\alpha$, and \ion{Fe}{xxvi}\,Ly$\alpha$ lines. Panel (c) shows $C$ residuals for the same best-fit continuum model without any emission lines.}
    \label{fig:hetg}
\end{figure}

\pp{Upon visual inspection of MEG and HEG spectra (see Fig.~\ref{fig:hetg}), we did not notice the presence of strong absorption or emission lines. We expect at least a significant iron line emission, as the analysis of the \nicer monitoring data consistently found Fe K emission lines with equivalent widths on the order of a few tens of eV. While the Fe K band emission in \nicer appears a broad feature, we expect to find the neutral and highly ionized lines well resolved in the grating spectra. In order to constrain limits on the expected emission lines, we added four narrow Gaussian lines at the energies of \ion{Si}{xiv}\,Ly$\alpha$, Fe\,K$\alpha$, \ion{Fe}{xxv}\,He$\alpha$, and \ion{Fe}{xxvi}\,Ly$\alpha$ (2.41\,keV, 6.40\,keV 6.70\,keV, and 6.97\,keV) with their width fixed at 10\,eV. Adding those features improves the fit to $C$-stat/d.o.f.\ of $510.80/422=1.21$, although the lines still appear very weak. The best-fit parameters are given in Table~\ref{tab:hetg_bestfit}.}  The most striking result of the \chandra observation is that the low Fe K line flux does not seem to be consistent with the \nicer observations. We note that the \nicer monitoring also indicated a moderate variability of the Fe line equivalent width and it is possible that the \chandra observation happened to catch an episode of very low line emission. Additionally, some of the observed Fe K line flux in \nicer may result from blended, unresolved, intermediate ionization states. Due to 
\chandra's lower collecting area, these enhanced emission lines will be more difficult to disentangle from the underlying continuum. We note that Fe K line emission in \nicer is best fit with a broad Gaussian with a centroid energy between neutral \ion{Fe}{i} and \ion{Fe}{xxv,} which supports the idea of a wide range of ionization states.

\begin{table}
 \centering
 \caption{Best-fit parameters for \chandra/HETG spectrum of \exo. All line widths were fixed to 10\,eV.}
 \label{tab:hetg_bestfit}
    \begin{tabular}{lc}
      \hline\hline
     parameter & value \\
     \hline
     $N_\mathrm{H}$\tablefootmark{a} & $2.26^{+0.06}_{-0.07}$ \\
     $F_\mathrm{PL}$\tablefootmark{b} & $4.49\pm0.03$ \\
     $\Gamma$ & $1.05^{+0.06}_{-0.05}$ \\
     $kT_\mathrm{BB}$\tablefootmark{c} & $0.63^{+0.04}_{-0.06}$ \\
     $F_\mathrm{BB}$\tablefootmark{b} & $0.28\pm0.02$ \\
     $E_\mathrm{{\ion{Si}{xiv}\,Ly}\alpha}$\tablefootmark{c} & $2.41^\dagger$\\
     $A_\mathrm{{\ion{Si}{xiv}\,Ly}\alpha}$\tablefootmark{d} & $0.74^{+0.29}_{-0.30}$\\
     $E_\mathrm{{Fe\,K}\alpha}$\tablefootmark{c} & $6.40^\dagger$\\
     $A_\mathrm{{Fe\,K}\alpha}$\tablefootmark{d} & $0.27^{+0.22}_{-0.18}$\\
     $E_\mathrm{{\ion{Fe}{xxv}\,He}\alpha}$\tablefootmark{c} & $6.70^\dagger$\\
     $A_\mathrm{{\ion{Fe}{xxv}\,He}\alpha}$\tablefootmark{d} & $0.42^{+0.25}_{-0.24}$\\
     $E_\mathrm{{\ion{Fe}{xxvi}\,Ly}\alpha}$\tablefootmark{c} & $6.97^\dagger$\\
     $A_\mathrm{{\ion{Fe}{xxvi}\,Ly}\alpha}$\tablefootmark{d} & $<0.38$\\
         \hline\hline
      \end{tabular}
        \tablefoot{
        \tablefoottext{a}{In $10^{22}\,\mathrm{cm}^{-2}$.}
       \tablefoottext{b}{Unabsorbed 0.1--10\,keV flux in $10^{-9}\,\mathrm{erg}\,\mathrm{s}^{-1}\,
        \mathrm{cm}^{-2}$.}
        \tablefoottext{c}{In keV.}
        \tablefoottext{d}{Photon flux in $10^{-3}\,\mathrm{phs}\,\mathrm{cm}^{-2}\,\mathrm{s}^{-1}$.}
           }
  \end{table}

Possible absorption features that are reported in the literature and might be associated with a CRSF are at ${\sim}10\,\mathrm{keV}$ \citep{Klochkov2007}, ${\sim}36\,\mathrm{keV}$ \citep{Reig1999} and ${\sim}63\,\mathrm{keV}$ \citep{Klochkov2008}. We therefore tested how much adding those absorption features improved the fit. We only found a slight improvement when including a 36\,keV line in Obs.~I, otherwise only upper limits on the line strength can be given (see Table~\ref{tab:crsf_test}). To test the presence of these lines, we kept the line widths fixed at conventional values (note that \citealt{Reig1999} used a Lorentzian and reported a width of 1\,keV of the 36\,keV feature), but we left the centroid energy free within a very narrow range since CRSF energies are known to vary with time and luminosity \citep{Staubert2019}. 


\begin{table}
\centering
\caption{Centroid optical depth $\tau$ of absorption features at previously reported CRSF energies. Line widths have been fixed during fitting. }
\label{tab:crsf_test}
   \begin{tabular}{cccc}
   \hline\hline
    energy & width & Obs.~I & Obs.~II\\
    \hline
    10\,keV & 2.5\,keV & $0.049^{+0.003}_{-0.004}$ & $0.038\pm0.008$ \\
    36\,keV & 1.0\,keV & $0.048\pm0.020$ & $\le0.025$\\
    63\,keV & 5.0\,keV &$\le0.025$ & $\le0.048$ \\
      \hline\hline
      \end{tabular}
  \end{table}

\section{Pulse-phase-resolved spectroscopy}\label{sec:phaseres}

\begin{figure}
    \centering
    \resizebox{\hsize}{!}{\includegraphics{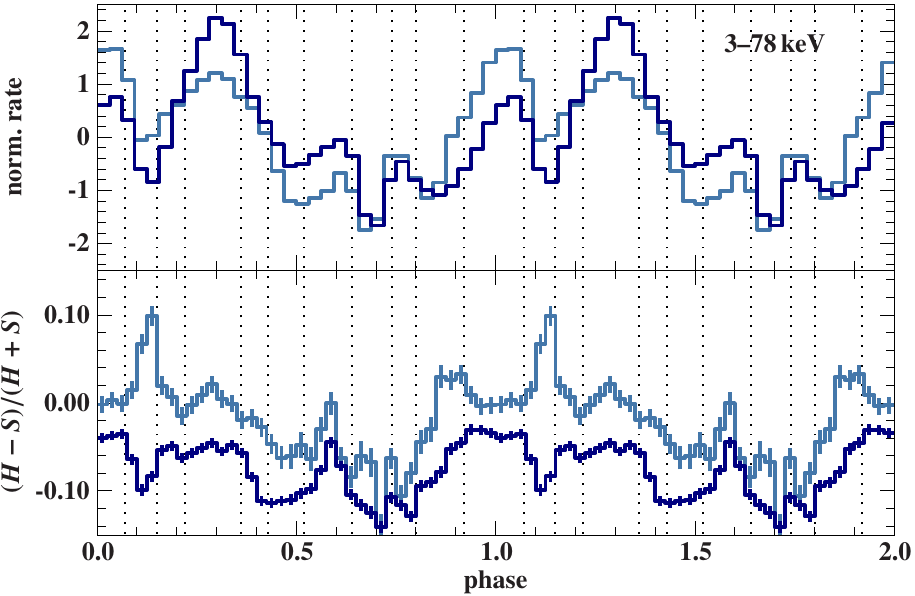}}
    \caption{Comparison of pulse profile and hardness ratio between Obs.~I and II. \textsl{Top:} \nustar broadband pulse profile for Obs.~I (dark blue) and Obs.~II (light blue). The pulse profiles are normalized to average zero and units of standard deviation. Dotted lines mark the phase intervals selected for pulse-phase-resolved spectroscopy. \textsl{Bottom:} Hardness ratio of the 3--7\,keV and 7--14\,keV band, defined as $(H-S)/(H+S)$. The phase definition is the same as in Paper I.}
    \label{fig:pp_obsI+II}
\end{figure}

\exo is known for the significant variability of its pulse profile with luminosity and the \nustar observations aimed at capturing broadband spectral transitions associated with the pulse profile evolution. The pulse profile evolution as observed with \nicer was presented in Paper I. We obtained the pulse period at the time of the \nustar observations by epoch-folding barycentered, dead-time-corrected light curves with a 0.5\,s binning. The resulting broadband pulse profiles and pulse-phase-resolved hardness ratios are shown in Fig.~\ref{fig:pp_obsI+II}. The pulse profile transition reported by \nicer and  \swift is characterized by a change in the relative intensity of the observed pulses around $\phi=1.0$ and $\phi=1.3$, respectively (as defined in Fig.~\ref{fig:pp_obsI+II} and Paper I). In addition to the sharp feature at $\phi{\sim}0.75$ reported in Paper I, the pulse-phase-resolved hardness ratio shows a dramatic change around the dip at $\phi=1.1\text{--}1.2$, which is softer than the neighboring peaks in Obs.~I, but harder in Obs.~II.

\begin{figure*}
    \resizebox{\hsize}{!}{\includegraphics{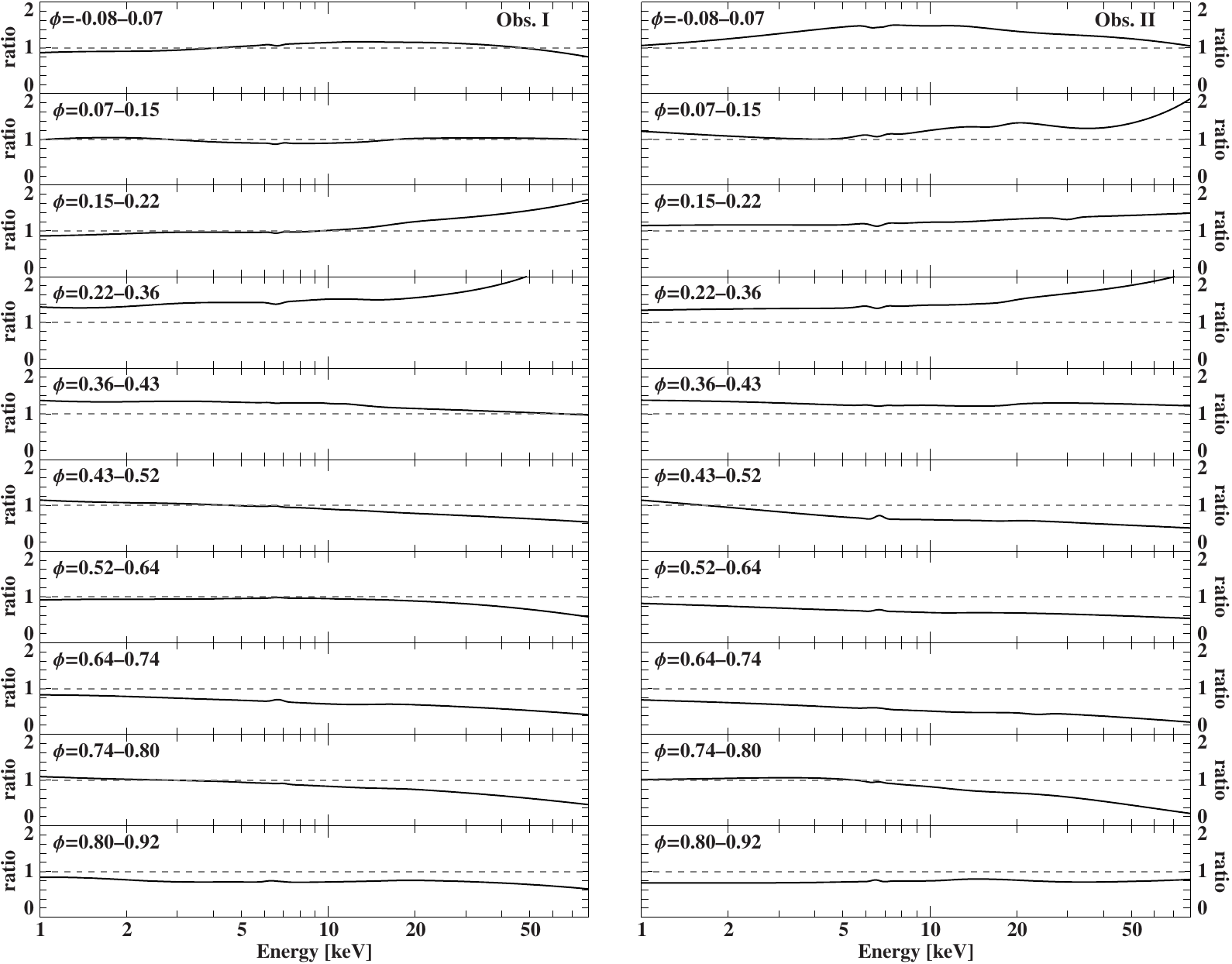}}
    \caption{Ratio plots of best-fit pulse-phase-resolved model to pulse-phase-averaged model.}
    \label{fig:phaseres_model}
\end{figure*}

The pulse-phase-resolved hardness ratios not only indicate a variation of spectral shape with luminosity, but also with the rotational phase of the neutron star.  The latter variation changes strikingly between the two observations. We therefore performed pulse-phase-resolved spectroscopy of the data of both \nustar observations to study the nature of this variability in more detail. Our selection of phase intervals (e.g., Fig.~\ref{fig:phaseres_model}) is aimed at capturing all significant changes in hardness ratios of several different energy bands while maintaining sufficient a signal-to-noise ratio in each phase interval to obtain meaningful parameter constraints.

The model used for fitting the pulse-phase-resolved spectra is the same as that used for the pulse-phase-averaged data, but we fixed the gain shift and \nustar/FPMB flux constant at the values found in Sect.~\ref{sec:phaseavg}. Our initial fits further showed a degeneracy between the absorption column and the black-body parameters, so we also fixed $N_\mathrm{H}$ to the pulse-phase-averaged values for each observation. As the 10\,keV feature in Obs.~II  is shallower and statistics are worse, we further fixed its width to the pulse-phase-averaged value. In our fits we  observed a slight variability of ${\sim}5\%$ of the \nicer/XTI flux constant relative to \nustar/FPMA with pulse phase. We confirmed that this behavior is a result of the much shorter \nicer/XTI exposures and reflects pulse-pulse variability between the simultaneous and non-simultaneous segments of the \nustar observation. These changes are at a level that does not affect our overall conclusions.
\begin{figure*}
    \centering
    \resizebox{\hsize}{!}{\includegraphics{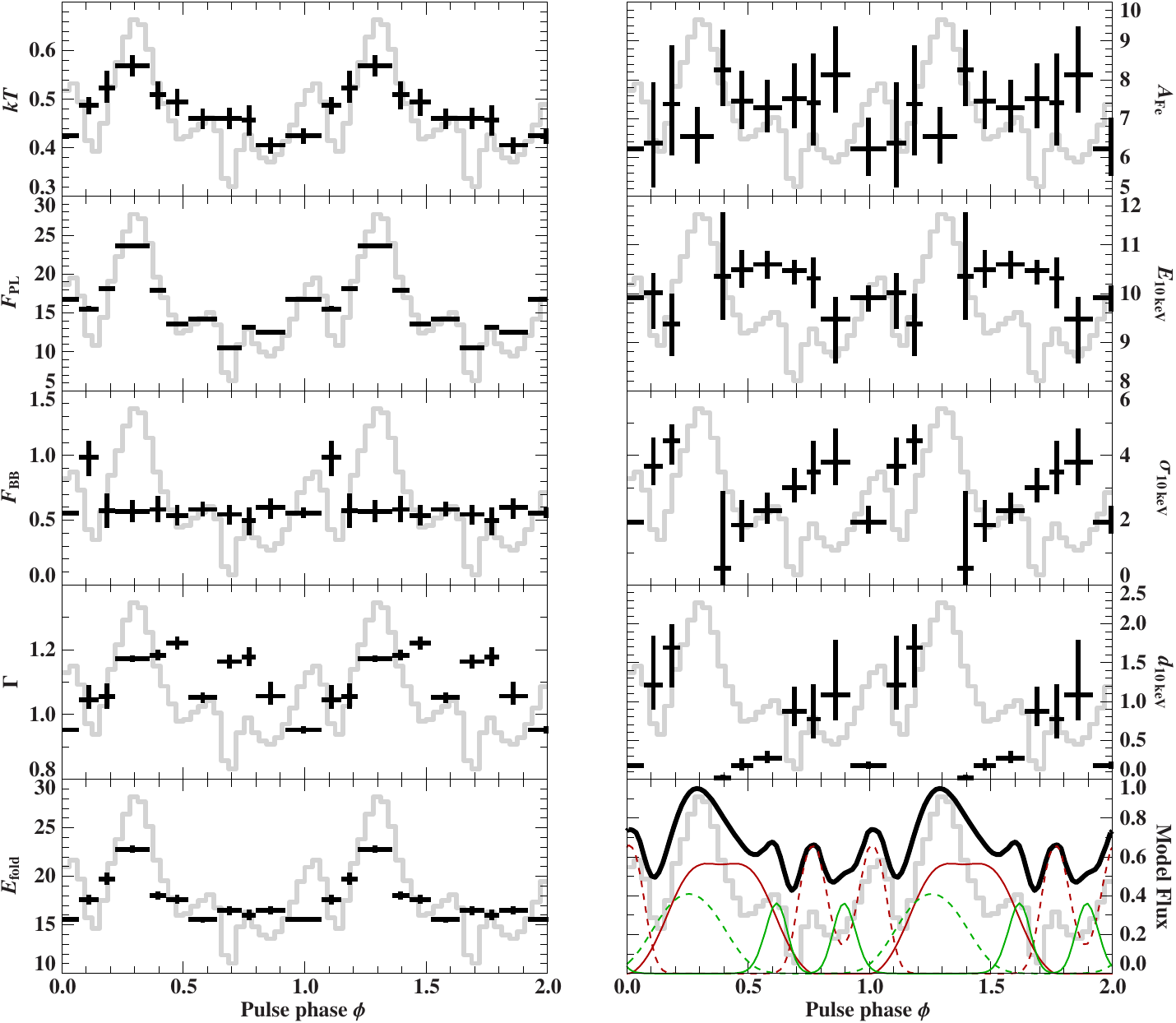}}
    \caption{Evolution of spectral model parameters with pulse phase
      (black) for Obs.~I, along with the component-resolved column emission as
      presented in Paper I. The model consists of an absorbed power law with exponential folding and an additional black-body component, as well as Fe\,K$\alpha$ emission line and an absorption feature with Gaussian optical depth (\texttt{gabs}). The free parameters (top to bottom, left to right) are the black-body temperature in keV; the power-law and black-body fluxes in $10^{-9}\,\mathrm{erg}\,\mathrm{s}^{-1}\,\mathrm{cm}^{-2}$; the photon index; the folding energy in keV; the Fe\,K$\alpha$ line flux in $10^{-3}\,\mathrm{photons}\,\mathrm{cm}^{-2}\,\mathrm{s}^{-1}$; the energy, width, and strength of the Gaussian absorption feature in keV. The depth of the 10\,keV absorption feature is consistent with zero at $\phi{\sim}0.3,$ and energy and width cannot be constrained. The bottom right panel shows the pulse profile model with a solid and dashed line
      indicating wall and top emission components, respectively, for the two
      accretion columns and the full pulse profile in black. All panels show the 3--78\,keV \nustar pulse profile of Obs.~I in light gray. The pulse profile model was obtained by fitting the 3.5--4.5\,keV \nicer pulse profile of ObsID~4201960113.}
    \label{fig:ppmod_o1}
\end{figure*}

\begin{figure*}
    \centering
    \resizebox{\hsize}{!}{\includegraphics{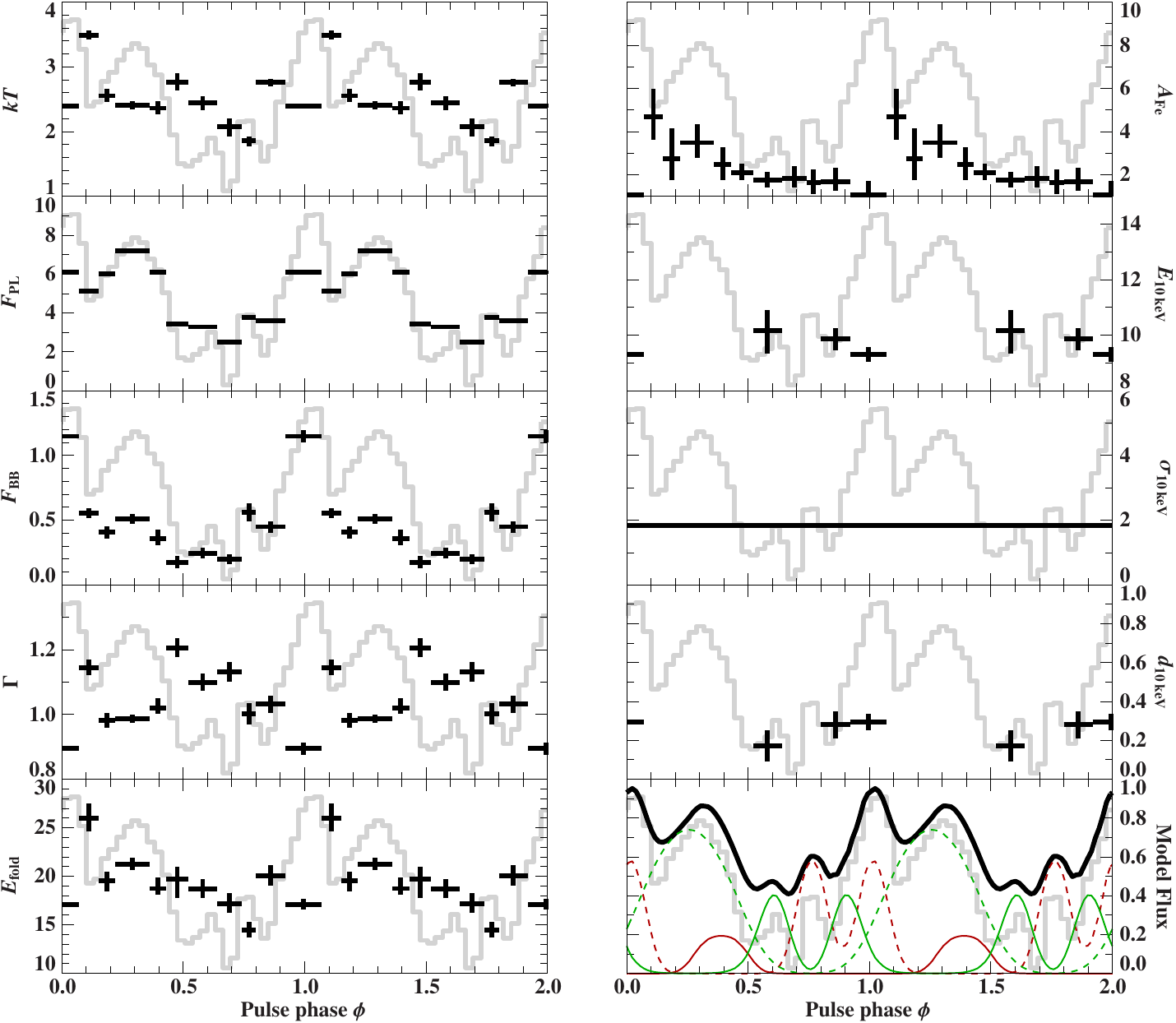}}
    \caption{Evolution of spectral model parameters with pulse phase (black) for Obs. II. Panels and units are the same as Fig.~\ref{fig:ppmod_o1}, but for Obs.~II. The width of the 10-keV absorption feature was fixed to the pulse-phase-averaged value of 1.8\,keV. Energies of 10-keV absorption feature are not shown where the depth is consistent with zero.  All panels show the 3--78\,keV \nustar pulse profile of Obs.~II in light gray.  The pulse profile model was obtained from fitting the 3.5--4.5\,keV \nicer pulse profile of ObsID~4201960150.} 
    \label{fig:ppmod_o2}
\end{figure*}

The parameter evolution with pulse phase for both observation is shown
in Figs.~\ref{fig:ppmod_o1} and~\ref{fig:ppmod_o2}. While the total flux difference between the two observations is about a factor of three, the flux of the black-body flux barely changes with luminosity. In Obs.~I, the black-body flux contribution is almost constant, except for the $\phi=0.07\text{--}0.15$ phase interval, where it roughly doubles compared to average. In Obs.~II, the black-body flux appears more variable, but it also peaks at a similar phase as in Obs.~I. The black-body temperature in Obs.~II is constantly higher than in Obs.~II, as expected from the pulse-phase-averaged fit.  In Obs.~I, the black-body temperature varies moderately and roughly correlates with the main peak of the pulse profile. In Obs.~II, the temperature evolution is more complex and also peaks at $\phi=0.07\text{--}0.15,$ where a spectral hardening is observed in Fig.~\ref{fig:pp_obsI+II}. We note that a systematic effect of fixing the photoelectric absorption component for the pulse-phase-resolved fits on the black-body temperature cannot be excluded, especially for Obs.~I, where the flux of this component peaks below 2\,keV.

The power-law continuum behaves similarly in both observations. Again, the most striking differences appear around phases $\phi=0.07\mbox{--}0.36$, where the folding energy and photon index differ substantially between the observations. Overall, the power-law continuum parameters evolve rather smoothly.

\section{Discussion}\label{sec:discussion}

\subsection{Continuum evolution with luminosity}

Assuming a distance of 2.4\,kpc, the inferred 3--50\,keV luminosities for the times of the two \nustar observations are ${\sim}1.1\times10^{37}\,\mathrm{erg}\,\mathrm{s}^{-1}$ and ${\sim}3.6\times10^{36}\,\mathrm{erg}\,\mathrm{s}^{-1}$, respectively. The luminosity is therefore comparable to the critical luminosity of  $(2.3\text{--}4.6)\times10^{36}\,\mathrm{erg}\,\mathrm{s}^{-1}$, which was derived from the 3--30\,keV flux, for the system, estimated by \citet{Epili2017} from comprehensive \rxte data of type~I and type~II outbursts of \exo\footnote{\citet{Epili2017} derived a critical luminosity of $(2\text{--}4)\times10^{37}\,\mathrm{erg}\,\mathrm{s}^{-1}$ for an assumed distance of 7.1\,kpc.}.  This is consistent with the results of \citeauthor{Epili2017}, which found the spectrum to harden as the luminosity decreases for the luminosity range of the \nustar observations.  We also note, however, that in Paper I (Fig.~5), the two \nustar observations clearly appear at the super-critical branch of the hardness-luminosity diagram. While there is a systematic shift in derived luminosity due to the different energy bands, it also reflects that the spectral hardening is not solely caused by a change in the power-law index or exponential cutoff, but rather due to a complex change in broadband spectral shape, which in our model is mostly reflected by an increase in the black-body temperature. Empirically, this hot black-body serves as a hard broad emission component. The power-law flux roughly follows the evolution of the overall luminosity.  \citet{Becker2007} propose a physically motivated spectral model of accretion-powered neutron stars based on solving the radiative transfer equation of the accretion column under certain assumptions. In their model, the seed photons form black-body and cyclotron emission, as well as bremsstrahlung, which experience thermal and bulk Comptonization in the infalling plasma. In this picture, the spectral transition from Obs.~I to~II  may be interpreted as a decrease of bremsstrahlung emission at a lower accretion rate. Accordingly, harder black-body emission is expected for higher impact velocities on the neutron star surface.

\citet{Postnov2015} studied the spectral hardness evolution of accreting pulsars with luminosity -including \exo- using \rxte/ASM data. They proposed a surface reflection model to explain the hardening with luminosity that tends to saturate at intermediate luminosities of what would be  few times $10^{36}\,\mathrm{erg}\,\mathrm{s}^{-1}$ for our assumed distance of 2.4\,kpc\footnote{In \citet{Postnov2015}, the saturation is reported above $10^{37}\,\mathrm{erg}\,\mathrm{s}^{-1,}$ but they also assume a distance of ${\sim}7\,\mathrm{kpc}$.}. Their observed hardness evolution using \rxte/ASM bands of 1.33--3\,keV and 5--12\,keV is in contrast with the hardness-ratio evolution observed by \citet{Epili2017} at the 3--10\,keV and 10--30\,keV bands that peaks at slightly lower luminosity. Accounting for the difference in assumed distance, we find our \nustar and \nicer hardness ratio evolution to be in agreement with the long term \rxte/PCA and \rxte/HEXTE motoring reported by \citet{Epili2017}.

The complex spectral shape of the soft continuum in Obs.~I cannot be modeled solely with a single absorbed power law and black body. We find that a partial covering absorber improves the description of the soft spectrum, but without the clear detection of absorption edges, the physical interpretation of the partial covering absorber cannot be robustly confirmed. At the available spectral resolution, physical properties of the absorbing medium (e.g., ionization), or an even more complex emission spectrum, cannot be properly constrained.

\subsection{Absorption features}

\exo is an actively debated CRSF candidate source with a number of potential spectral features. The first suggestion of a CRSF detection at ${\sim}36\,\mathrm{keV}$ was reported by \citet{Reig1999}, who observed an absorption-like feature in  stacked \rxte/HEXTE observations. Due to the known variability, the complex continuum shape, and the absence of a harmonic feature, \citet{Reig1999} cautioned, however, that a definitive statement on the nature of the 36\,keV  feature cannot be made. In contrast to these earlier analyses, in their reanalysis of the \rxte data \citet{Epili2017} did not find any indications for absorption features that could be potential CRSFs in \rxte data.

Another possible absorption feature at ${\sim}63\,\mathrm{keV}$ was detected by \citet{Klochkov2008} in archival \integral data of the 2006 giant outburst. \citet{Klochkov2008} were only able to detect this feature during a relatively short pulse-phase range and not in the pulse-phase-averaged spectrum. The data quality makes the assessment of the statistical significance even more difficult. None of the authors who analyzed the \rxte/HEXTE data of \exo report the presence of such a feature.

We tested for the presence of these absorption features, but we cannot confirm the presence of either the 36\,kev or 63\,keV lines. Adding these lines only leads to a minor improvement of the fit statistics for the 36\,keV in Obs.~I. We note, however, that only an upper limit for the optical depth can be obtained for Obs.~II, which, albeit fainter, still has excellent statistical quality. We further note that line energies and widths had to be fixed during the fit in order to obtain meaningful results for the optical depth. Support for these claimed CRSF detections from these two \nustar observations therefore remains tenuous at best.

The most robustly detected absorption-like feature discussed in the literature appears around ${\sim}10\,\mathrm{keV}$. It was first reported by \citet{Klochkov2007} in \integral spectra and later confirmed by \citet{Wilson2008} using \rxte; both concerned the 2006 giant outburst. \nustar observed the 2021 giant outburst twice, and \citet{Tamang2022} report the detection of a ${\sim}10\,\mathrm{keV}$ absorption feature in Obs.~I, which is in line with our detection in both \nustar observations, albeit at smaller optical depth in Obs.~II. While we can therefore confirm the presence of an $\sim$10\,keV absorption-like feature, its interpretation as a CRSF remains unconfirmed. 

The most prominent CRSF source to compare with is evidently 4U\,0115$+$63, which exhibits a fundamental CRSF at ${\sim}11\,\mathrm{keV}$ \citep[e.g.,][and references therein]{Heindl1999, Mueller2013, Staubert2019, Kuehnel2020} and despite its complex continuum shape and variability allows for the firm detection of up to five harmonics. 
While the depth and width of absorption feature in the spectrum of \exo are comparable with those of the fundamental CRSF in 4U\,0115$+$63, we do not find signatures of absorption features at harmonic energies of the ${\sim}10\,\mathrm{keV}$ feature. Despite the many unknowns in the physical details of CRSF formation, we are not aware of a strong argument that prevent harmonic lines from forming for $B$-fields as low as an ${\sim}10\,\mathrm{keV}$ fundamental \citep[see, e.g.,][]{Schoenherr2007}. It is therefore entirely possible that the observed absorption dip is the result of the purely empirical continuum modeling and not physical. We emphasize that similar line-like features that are not caused directly by the cyclotron resonance have been observed in a number of sources, although generally at much lower luminosity, where the spectral transition toward a two-component spectrum mimics an absorption feature between 10\,keV and 20\,keV on top of smooth power-law continuum. Examples include GX\,304$-$1 \citep{Tsygankov2019a}, A\,0535$+$26 \citep{Tsygankov2019b}, GRO\,J1008$-$57 \citep{Lutovinov2021}, and SRGA\,J124404.1$-$632232 \citep{Doroshenko2022}. We emphasize that this behavior has typically been seen at luminosities far below those of the \nustar observations of \exo, but the spectral hardening and the failure of a simple cutoff power-law model in Obs.~II indicate a possible over-simplification of the continuum modeling.

For low-luminosity states, the complex shape of the continuum is explained by the spectral formation in an atmosphere of the neutron star, strongly overheated by direct stopping via Coulomb collisions \citep{Sokolova-Lapa2021, Mushtukov2021}. An intense redistribution near the fundamental cyclotron resonance at electron temperatures of ${\sim}10$--$40\,\mathrm{keV}$ results in the formation of a strong red wing of the CRSF \citep[the effect, although less pronounced, is also well known for electron temperatures of ${\sim}5\,\mathrm{keV}$, which are typically expected in accretion columns;][]{Schwarm2017b, Schoenherr2007}. The transition from low-luminosity Coulomb braking to the matter stopping at higher luminosities is not well studied. However, in the case of a gradual transition, one can expect a similar but weaker mechanism in play at intermediate luminosities. A lack of any clear indication of a cyclotron resonance at energies above $30$--$50\,\mathrm{keV}$ makes this interpretation doubtful.

Polarization effects in the strong magnetic field can also result in a complex shape of the Comptonized continuum due to the contribution of the photons of two polarization modes, which have principally different properties in interactions with electrons \citep{Meszaros1985b}. The two modes have different spectral dependencies and a ratio of their intensities changes across the broadband continuum, which can result in local excesses or shallow dips in a total spectrum, summed over the polarization modes. In addition to a classical magnetized plasma, the vacuum polarization effect influences the spectral formation in the strong magnetic field \citep{Adler1971, Meszaros1978}. This effect has not been investigated in greater detail for $B{\sim}10^{12}\,\mathrm{G}$, but recent advances in the field show that it is capable of enhancing the mode interchange across the spectrum and can also result in the formation of local spectral suppressions and shallow dips due to the vacuum resonance \citep{Sokolova-Lapa2023}.

\subsection{Spectral evolution with pulse-phase and emission geometry} 

In addition to the high luminosity observations taken by various X-ray missions during the recent and historic giant outbursts, \xmm and \nustar observed \exo in a low luminosity state; \xmm at 
the onset of a type~I outburst \citep{Ferrigno2016} and \nustar \citep{Fuerst2017} and \astrosat \citep{Jaisawal2021} close to the propeller regime. These low luminosity observations are particularly interesting as \exo exhibited a sudden change in the photon index around $\phi\sim0.65$\footnote{\citet{Fuerst2017} reported this feature at phase $\phi\sim0.85$ and \citet{Ferrigno2016} at $\phi\sim0.27$ according to their phase definitions. Phase-matching is not entirely unambiguous due to the different pulse profile morphology at low luminosity and was done by extrapolation of the luminosity-resolved pulse profiles in \citet{Thalhammer2022} and \citet{Epili2017}.}. While our pulse-phase binning is most likely too coarse to resolve this feature, a similarly narrow drop can be seen in the hardness ratio shown in Fig.~\ref{fig:pp_obsI+II}.

Our interpretation of the variation of the spectral parameters in terms of the emission of the accretion column is based on the relativistic ray-tracing approach discussed in further detail in Paper I. We are able to reproduce the \nicer pulse profiles associated with the two \nustar observations using a two-column model including identical columns of 2\,km in height and 300\,m in radius that emit both from the walls (fan beam) and top (pencil beam); these are located near the equator of the neutron star. In this model, the observed asymmetry of the pulse profiles with respect to rotational phase is due to the two columns having different emission patterns, as well as a slight displacement of the columns compared to a purely dipolar $B$-field. The evolution of the relative strengths of the peaks of the pulse profiles from one observation to the other is explained solely by a change of the emission patterns of the wall and top of each column, as well as their flux normalizations. These changes lead to an apparent phase shift and reduction of the width of the individual sub-peaks, which when superposed cause a strong change in the total pulse profile. The model of Paper I yields a relative decomposition of the total pulse profile into fan- and pencil-beam contributions of each column, so we can attempt an association of the spectral variability with the visibility of these individual components. We caution in Paper I that the full parameter space cannot be explored sufficiently to exclude other possible solutions. Uncertainty estimates on parameters are rather crude as the behavior of a Markov chain Monte Carlo (MCMC) analysis is dominated by systematic rather than statistical uncertainties. The decomposed pulse profile is shown  in Figs.~\ref{fig:ppmod_o1} and \ref{fig:ppmod_o2}. A clear association of spectral components with emission components remains inconclusive. This is likely a result of the empirical spectral model and the oversimplified decomposition of the complex X-ray spectrum into a power-law and a black-body component. 

To our knowledge, the only currently available radiative-transfer models for accretion columns that makes predictions of the spectral energy distribution of the radiation  escaping the top of the column are those of \citet[][hereafter WWB17]{West2017a, West2017b} and of \citeauthor{Becker2022} (\citeyear{Becker2022}; hereafter BW22). These authors find that in high-luminosity sources the predominant contribution to the total X-ray emission emerges from the column walls. A transition to a strong column top contribution is only expected at low mass-accretion rates, far below even the fainter \nustar observation. This result goes against the pulse-profile decomposition, which requires emission from the sides and top of the column.

As an alternative model, \citet{Postnov2015} took the reflection of the relativistically downward-beamed wall emission off the neutron star surface into account; this could lead to a similar emission pattern as the traditional cap or pencil component. However, in their analysis of a sample of accreting X-ray pulsars (including \exo), the reflected component contributes substantially to a spectral hardening. Its contribution grows with increasing luminosity, which implies a hardness ratio evolution with luminosity that is opposite what we observe.

\section{Conclusions and outlook}\label{sec:outlook}

In this work, we present the broadband spectral analysis of two \nustar observations taken during the giant 2021 outburst of \exo, once near the peak and once during the decline phase of the outburst. We find a drastic spectral transition characterized by a spectral hardening toward lower luminosity. This hardening cannot be described by a simple change in the power-law index or folding energy; it requires a more complex modification of the continuum model. This behavior is somewhat surprising as many accreting pulsars show stable, power-law-like continuum mostly formed by Comptonized bremsstrahlung at luminosities above $10^{37}\,\mathrm{erg}\,\mathrm{s}^{-1,}$ and significant spectral transitions toward low-luminosity accretion are expected at luminosities at least one order of magnitude lower. Detailed radiative transfer calculations with the available column models are hampered by the unknown magnetic field strength. The only absorption feature that is robustly detected in both observations is the previously reported ${\sim}10\,\mathrm{keV}$ feature. A magnetic origin is, however, highly questionable because of the lack of harmonic features, which leads us to believe that this feature is a product of the complex continuum formation. The $B$-field estimate in \exo therefore remains an open question.

The pulse-profile model presented in Paper~I offers a possible emission geometry including a column wall and top component. Our attempt to associate the spectral components with these wall and top components remains somewhat inconclusive and reveals the limited physical interpretability of the traditional phenomenological spectral models. The \nustar observations and the pulse profile models raise, however, the intriguing question as to how the emission from the top of the column, which is strongly required by most pulse profile models \citep[for another example, see ][]{Iwakiri2019}, manifests itself in the observed X-ray spectrum. While the radiative transfer models WWB17 and BW22 include a top emission component, its contribution is supposed to be marginal at high luminosities. \citet{Postnov2015} suggested reflected radiation from the neutron star surface as an explanation for some of the luminosity dependence of the broadband X-ray spectrum, but the behavior predicted by this model is opposite to what is seen here. A solution to these problems might be that the accretion geometry of \exo is more complicated than a simple skew-symmetric dipole. Such an increased accretion complexity is proposed by \citet{Malacaria2023}, based on the small polarization degree of \exo that was observed by \textsl{IXPE}; it is also supported by timing data of \textsl{Insight}-HXMT \citep{Fu2023}.

We expect major progress in the development of ray-tracing pulse-profile models; this should include reflected radiation, obscuration by the column itself, or a hollow-accretion column, which will advance our understanding of the accretion geometries in \exo and other X-ray pulsars. While available physically motivated spectral models are very successful in describing the pulse-phase-averaged continuum emission at different luminosities, usable models for pulse-phase-resolved emission are not yet widely available. The deficient association of the empirical spectral evolution with the geometrical emission components underlines the need for active development in this direction.  An attempt to model the pulse-phase-averaged broadband X-ray spectrum of \exo with self-consistent hydro dynamical models such as WWB17 and a detailed study of the energy-resolved pulse profiles are beyond the scope of this work and will be presented in a forthcoming publication.

\begin{acknowledgements}
We thank the \nustar, \swift, and \chandra PIs and teams for approving our DDT requests and the effort of scheduling this dense monitoring campaign. We further thank Norbert Schulz for insightful discussions of the \chandra/HETG spectrum. We thank the anonymous referee for useful comments and suggestions that helped to improved the manuscript.  RB acknowledges support by NASA under award number 80NSSC22K0122. The material is based upon work supported by NASA under award number 80GSFC21M0002. ESL\ and JW\ acknowledge partial funding under Deutsche Forschungsgemeinschaft grant WI 1860/11-2 and Deutsches Zentrum f\"ur Luft- und Raumfahrt grant 50 QR 2202. Portions of this work performed at Naval Research Laboratory were supported by NASA. This research has made use of data from the \nustar mission, a project led by the California Institute of Technology, managed by the Jet Propulsion Laboratory, and funded by the National Aeronautics and Space Administration. Data analysis was performed using the \nustar Data Analysis Software (NuSTARDAS), jointly developed by the ASI Science Data Center (SSDC, Italy) and the California Institute of Technology (USA). This research has further made use of data obtained from the Chandra Data Archive and the Chandra Source Catalog, and software provided by the Chandra X-ray Center (CXC) in the application package CIAO.
\end{acknowledgements}

\end{document}